\documentclass[]{article}

\input glyphtounicode
\usepackage[T1]{fontenc}
\usepackage[utf8]{inputenc}

\usepackage{ifdraft}

\usepackage[english]{babel}
\usepackage{amsmath}
\usepackage{mathtools}
\usepackage{enumitem}
\usepackage[amsmath,thmmarks,hyperref]{ntheorem}
\usepackage{siunitx}
\usepackage{csquotes}

\usepackage{xcolor}
\definecolor{hypercolor}{rgb}{0,0.2,0.7}

\usepackage{footmisc}

\usepackage[numbers,sort&compress]{natbib}
\usepackage[final]{graphicx}
\usepackage{tikz}
\usetikzlibrary{calc,decorations.markings}

\newif\iffancyfont%
\fancyfontfalse%

\IfFileExists{MinionPro.sty}{\IfFileExists{MyriadPro.sty}{\fancyfonttrue}{}}{}

\iffancyfont%
  \usepackage[lf,medfamily,italicgreek,openg]{MinionPro}
  \usepackage[lf,medfamily,onlytext,scale=0.96]{MyriadPro}

  \DeclareMathSymbol{\widehatsym}{\mathord}{largesymbols}{'302}

  \usepackage[toc,eqno,bib]{tabfigures}
  \newtagform{tabular}[\figureversion{tabular}]{(}{)}
  \usetagform{tabular}
\else%
  \usepackage[charter,greekuppercase=italicized,cal=cmcal]{mathdesign}

  \DeclareMathSymbol{\widehatsym}{\mathord}{largesymbols}{"62}

\fi%

\usepackage{dsfont}

\usepackage[scr=rsfso]{mathalfa}

\ifdraft{

}{
  \usepackage{scalerel}
  \usepackage{accents}
  \newcommand\lowerwidehatsym{%
    \text{\smash{\kern-.1ex\raisebox{-1.3ex}{%
      $\widehatsym$}}}}

}

\usepackage{microtype}

\usepackage[sf,bf,small]{titlesec}
\usepackage[labelfont={sf,bf},labelsep=period]{caption}
\usepackage{subcaption}

\setlist[1]{labelindent=\parindent}
\setlist[description]{font=\sffamily\bfseries,align=right,labelsep=1em}

\numberwithin{equation}{section}

\makeatletter
\newcommand{\subtitle}[1]{\newcommand{\@subtitle}{#1}}
\newcounter{and}
\newdimen{\instindent}
\newcommand{\institute}[1]{\newcommand{\@institute}{#1}}
\newcommand{\inst}[1]{\unskip\smash{$^{#1}$}\setcounter{and}{1}\ignorespaces}
\newcommand{\email}[1]{\href{mailto:#1}{#1}}
\renewcommand{\maketitle}{
  {
    \raggedright%
    \LARGE%
    \noindent%
    \bfseries%
    \sffamily%
    \@title%
    \newline
    \Large
    \@subtitle
    \par
  }

  \vspace{1.5\baselineskip}

  {
    \raggedright%
    \renewcommand{\and}{\unskip, \ignorespaces}%
    \noindent\ignorespaces\@author\par
  }

  \vspace{0.5\baselineskip}

  {
    \small%
    \parindent=0pt%
    \parskip=0pt%
    \setcounter{and}{1}%
    \renewcommand{\and}{%
      \par\stepcounter{and}%
      \hangindent\instindent%
      \noindent%
      \hbox to \instindent{\hss\smash{$^{\theand}$\enspace}}\ignorespaces%
    }%
    \setbox0=\vbox{\@institute}%
    \ifnum\value{and}>9\relax\setbox0=\hbox{$^{88}$\enspace}%
    \else\setbox0=\hbox{$^{8}$\enspace}\fi%
    \instindent=\wd0\relax%
    \ifnum\value{and}=1\relax%
    \else%
      \setcounter{and}{1}%
      \hangindent\instindent%
      \noindent%
      \hbox to \instindent{\hss\smash{$^{\theand}$}\enspace}\ignorespaces%
    \fi%
    \ignorespaces%
    \@institute\par
  }
}
\makeatother

\renewenvironment{abstract}{
  \addvspace{1.5\baselineskip}%
  \topsep=0pt\partopsep=0pt%
  \trivlist\item[\hspace{\labelsep}\bfseries\sffamily Abstract.]
}{}

\newenvironment{acknowledgments}{
  \addvspace{1.5\baselineskip}%
  \topsep=0pt\partopsep=0pt%
  \trivlist\item[\hspace{\labelsep}\bfseries\sffamily Acknowledgments.]
}{}

\definecolor{lime}{HTML}{A6CE39} 
\newcommand{\orcidicon}{%
	\begin{tikzpicture}
	\draw[lime, fill=lime] (0,0) 
	circle [radius=0.16] 
	node[white] {{\fontfamily{qag}\selectfont \tiny ID}};
	\draw[white, fill=white] (-0.0625,0.095) 
	circle [radius=0.007];
	\end{tikzpicture}
	\hspace{-3mm}
}
\newcommand\orcidSeSc{{\href{https://orcid.org/0000-0003-1997-0026}{\orcidicon}}}
\newcommand\orcidChPf{{\href{https://orcid.org/0000-0002-1712-6860}{\orcidicon}}}

\newcommand\latinabbr\textit

\newcommand{\dif}{\operatorname{d}\!}

\DeclarePairedDelimiterX\ip[2]{\langle}{\rangle}{#1 \,\delimsize\vert\, #2}

\theoremnumbering{arabic}
\qedsymbol{\ensuremath{\quad\Box}}

\theoremstyle{plain}
\theorembodyfont{\itshape}
\theoremheaderfont{\normalfont\sffamily\bfseries}
\theoremseparator{.}

\theorembodyfont{\upshape}

\theoremstyle{nonumberplain}
\theorembodyfont{\normalfont}
\theoremheaderfont{\normalfont\sffamily\bfseries}
\theoremsymbol{\ensuremath{\quad\Box}}

\title{\texorpdfstring{Static spherically symmetric black holes in weak $f(\mathbb{T})$-gravity}{Static spherically symmetric black holes in weak f(T)-gravity}}
\subtitle{}

\author{
  Christian Pfeifer\inst{1} \orcidChPf{}
  \and
  Sebastian Schuster\inst{2} \orcidSeSc{}
}

\institute{
  Center Of Applied Space Technology And Microgravity - ZARM, University of Bremen, Am~Fallturm~2, 28359 Bremen, Germany.
  E-mail:~\email{christian.pfeifer@zarm.uni-bremen.de}
  \and
  Ústav teoretické fyziky, Matematicko-fyzikální fakulta, Univerzita Karlova, V~Holešovičkách~2, 180~00~Praha~8, Czech Republic.
  E-mail:~\email{sebastian.schuster@utf.mff.cuni.cz}
}

\usepackage[breaklinks,final]{hyperref}
\hypersetup{
  unicode,
  colorlinks,
  linkcolor=hypercolor,
  urlcolor=hypercolor,
  citecolor=hypercolor,
  bookmarksnumbered,
  bookmarksdepth=2,
  pdfauthor={Christian Pfeifer, Sebastian Schuster},
  pdftitle={Surface Gravity of Spherical Black Holes in f(T)-Gravity}
}

\begin{document}

\maketitle

\begin{abstract}
    With the advent of gravitational wave astronomy and first pictures of the \enquote{shadow} of the central black hole of our milky way, theoretical analyses of black holes (and compact objects mimicking them sufficiently closely) have become more important than ever. The near future promises more and more detailed information about the observable black holes and black hole candidates. This information could lead to important advances on constraints on or evidence for modifications of general relativity. More precisely, we are studying the influence of weak teleparallel perturbations on general relativistic vacuum spacetime geometries in spherical symmetry. We find the most general family of spherically symmetric, static vacuum solutions of the theory, which are candidates for describing teleparallel black holes which emerge as perturbations to the Schwarzschild black hole. We compare our findings to results on black hole or static, spherically symmetric solutions in teleparallel gravity discussed in the literature, by comparing the predictions for classical observables such as the photon sphere, the perihelion shift, the light deflection, and the Shapiro delay. On the basis of these observables, we demonstrate that among the solutions we found, there exist spacetime geometries that lead to much weaker bounds on teleparallel gravity than those found earlier. Finally, we move on to a discussion of how the teleparallel perturbations influence the Hawking evaporation in these spacetimes.
\end{abstract}

\tableofcontents

%
\section{Introduction}
When studying isolated, astrophysical objects like stars, neutron stars, or black holes, the real physical system requires a high degree of sophistication and model building that can usually only be dealt with numerically. Nevertheless, simpler models (numerical or analytic) often provide the necessary stepping stones. Spherically symmetric solutions in particular are one of the simplest building blocks in this endeavour. From these solutions one obtains a first approximation of the motion of test particle around these objects, which lead to observables like the perihelion shift, light deflection or the Shapiro delay. A more realistic description of these objects usually requires axially symmetric solutions, to take their rotation into account.

In general relativity (GR) the Birkhoff theorem states that the unique spherically symmetric vacuum solution of the theory is the famous Schwarzschild solution. It is static, asymptotically flat and contains a black hole with an event horizon at the Schwarzschild radius \cite{Jebsen}. In modified theories of gravity, in general, the Birkhoff theorem in this strong form does not hold and the spherically symmetric vacuum solutions have to be analyzed in great detail: they are usually neither the Schwarzschild spacetime, nor static, nor unique. Weaker version of the Birkhoff theorem have been discussed in the context of $f(R)$-theory \cite{Capozziello:2011wg,Nzioki:2013lca}, scalar tensor theory \cite{Krori_1977} and $f(\mathbb{T})$-gravity \cite{Dong:2012en}, and further modified theories of gravity~\cite{Dai:2008zza}.

Among the many modifications and extensions of general relativity \cite{Nojiri:2006ri}, one famous and extensively studied one is Teleparallel gravity \cite{TPGravBook,Krssak:2018ywd,DeAndrade:2000sf}. It uses a tetrad and a flat, metric compatible spin connection with torsion, the so called Weitzenböck (or teleparallel) geometry \cite{Weitzenbock1923}---instead of pseudo-Riemannian geometry as defined by a metric and its metric-compatible, torsion-free Levi-Civita connection---to describe the dynamics of gravity. In this work we will construct a new family of static, asymptotically flat\footnote{In the context of teleparallel gravity asymptotically flat means that the metric obtained from the tetrad defines an asymptotically flat pseudo-Riemannian geometry.} spherically symmetric vacuum solutions to weak $f(\mathbb{T})$-gravity, which can be interpreted as teleparallel perturbations of the Schwarzschild black hole of general relativity. This family of solutions is not the unique family of asymptotically flat, spherically symmetric vacuum solutions of the theory, as different families of solutions with this property have been derived earlier. This finding explicitly demonstrates the failure of the Birkhoff theorem in $f(\mathbb{T})$-gravity. Having found weak $f(\mathbb{T})$-gravity black hole solutions we analyze their classical and semi-classical properties.

A starting point for the study of teleparallel theories of gravity is the reformulation of GR in the framework of teleparallel geometry called the "teleparallel equivalent of general relativity" (TEGR) \cite{TPGravBook,Maluf:2013gaa,Garecki:2010jj,BeltranJimenez:2019tjy}. TEGR is a theory that is dynamically fully equivalent to GR, but derived from an action involving the so called torsion scalar $\mathbb{T}$. It differs from the Einstein--Hilbert action by a boundary term $B$. Starting from this reformulation many teleparallel modifications of GR have been constructed \cite{Bahamonde:2015zma}, such as new general relativity \cite{Hayashi}, Born--Infeld gravity \cite{Ferraro:2008ey}, teleparallel Horndeski gravity \cite{Bahamonde:2019shr}, scalar torsion gravity \cite{Hohmann:2018vle,Hohmann:2018dqh,Hohmann:2018ijr}, and, in analogy to $f(R)$-theories \cite{Sotiriou:2008rp,DeFelice:2010aj}, $f(\mathbb{T})$-theories of gravity \cite{Ferraro:2006jd,Krssak:2015oua}. The latter theories have been studied in particular detail in regard on their consequences in cosmology \cite{Ferraro:2006jd,Bengochea:2008gz,Bamba:2010wb,Dent:2011zz,Cai:2015emx,Junior:2015bva} and astrophysics \cite{Cai:2018rzd,Ahmed:2016cuy,Hohmann:2018jso}, as well as their degrees of freedom \cite{Blixt:2020ekl,Jimenez:2020ofm}. Further generalizations have also been considered in the literature, including ones explicitly involving the boundary term in $f(\mathbb{T},B)$-gravity \cite{Bahamonde:2019jkf}, or, involving three terms, $\mathbb{T}_{\text{ax}}, \mathbb{T}_{\text{vec}}, \mathbb{T}_{\text{ten}}$; these are a specific decomposition of the torsion scalar $\mathbb{T}$, which feature in so-called $f(\mathbb{T}_{\text{ax}}, \mathbb{T}_{\text{vec}}, \mathbb{T}_{\text{ten}})$-gravity \cite{Bahamonde:2017wwk}.

The study of static spherically symmetric solutions in $f(\mathbb{T})$-gravity, and its $f(\mathbb{T},B)$ and $f(\mathbb{T}_{\text{ax}}, \mathbb{T}_{\text{vec}}, \mathbb{T}_{\text{ten}})$ generalizations \cite{Bahamonde:2020bbc,Bahamonde:2020vpb}, is an ongoing field of research \cite{Golovnev:2020las,Golovnev:2021htv}.

In $f(\mathbb{T})$-gravity, only few non-vacuum solutions are known. These usually are encountered in the context of anisotropic matter and Boson stars \cite{Daouda:2012nj,Horvat:2014xwa,Ilijic:2020vzu}. Also, studies of solutions sourced by a non-linear electromagnetic field exist \cite{Junior:2015fya}. Among the vacuum solutions are regular BTZ black hole solutions in Born--Infeld gravity \cite{Boehmer:2020hkn,Boehmer:2019uxv}, and teleparallel perturbations of Schwarzschild geometry in weak $f(\mathbb{T}) = \mathbb{T} + \frac{\alpha}{2} \mathbb{T}^2$ gravity \cite{DeBenedictis:2016aze,Bahamonde:2019zea}.

We revisit here these weak teleparallel perturbations of Schwarzschild geometry. They turn out to be parametrized by two constants of integration, whose value influences physically important properties of spacetime. In the previous studies the two constants of integration have been chosen such that the weakly teleparallel solution is asymptotically flat and---at large distance away from the central mass---is close to Schwarzschild geometry. As we will show, it turns out that these solutions are not the most general black hole solutions. More importantly, they are not the ones closest to Schwarzschild geometry. Instead of determining the constants of integration far away from the central mass, we fix one of the constants of integration by the value of the determinant of the metric at the horizon. Moreover, we demonstrate that the second integration constant can be chosen in such a way that the $rr$-component of the metric has the same fall-off property at large distance away from the central mass as in Schwarzschild geometry. This implies a similar behaviour for the $tt$-component and turns out to be sufficient for the spacetime to be asymptotically flat.

Thus, the solutions we present here, are the  general static, asymptotically flat black hole vacuum solutions of weak $f(\mathbb{T})$-gravity. They are parametrized by the value of the product of the determinant of the metric at the horizon. They contain the solutions found in \cite{DeBenedictis:2016aze,Bahamonde:2019zea} for a very specific value of the determinant of the metric at the horizon.\footnote{Unlike in the GR case of the Schwarzschild geometry, this determinant at the horizon is \emph{not} $- r^2 \sin\theta$.} These findings demonstrate that there is no unique spherically symmetric, asymptotically flat, static vacuum solution in this theory: The Birkhoff theorem does not hold.

Most observations in relativistic astrophysics rely on the perspective of general relativity. Therefore, it is important to consider how a metric gained from a different theoretic background appears through this lens. While our solution is a vacuum solution in weak $f(\mathbb{T})$ gravity, in light of Birkhoff's theorem, this cannot be the case if interpreted in strictly general relativistic terms. Rather, here it would appear to arise from an effective matter distribution. The corresponding, effective energy-momentum tensor is then entirely prescribed by the torsion tensor. It is important to point out that this is an analogy only---but this analogy can guide our expectation about the theory whence the metric originated.

Having identified the most general black hole solutions of weak $f(\mathbb{T})$-gravity, we analyze classical observables obtained from point particle motion: the photon sphere, the perihelion shift, the Shapiro delay, and the light deflection each to lowest order in the teleparallel perturbation. These observables can then be used to constrain the value of the coupling to the teleparallel perturbation based on observations. Interestingly, for the solutions we discuss here the constraints are weaker than the ones obtained for the previously found solutions. 

Since our solution family is still static, we can treat the putative horizon appearing as a Killing horizon. This grants us access to the zeroth law of black hole thermodynamics through its surface gravity \cite{WaldRacz1996}. For this we will further discuss the location of the horizon and its surface gravity. The Hawking effect being a kinematic effect \cite{EssIness}, this in turn gives us a first look at the semi-classical phenomenology of this new family of black holes. We will then use this window to study how this Hawking radiation would differ from that of a Schwarzschild black hole---based on the heuristic measure of \enquote{sparsity $\eta$}. This quantity provides a quick way to estimate the average density of state of the particles emitted by comparing their localization timescale to their emission rate. In other words, it is an estimator for how \enquote{classical} or \enquote{quantum} the produced radiation is.

The presentation of our results is structured as follows: Section~\ref{sec:fTBH} on black holes on weak $f(\mathbb{T})$-gravity we begin by recalling the main mathematical notions of $f(\mathbb{T})$-gravity in Section~\ref{ssec:covfT}. Then we investigate the general static, spherically symmetric solutions of weak $f(\mathbb{T})$-gravity and identify the most general black hole solutions among them in Section~\ref{ssec:stspBHs}. The short section~\ref{sec:gr} will then explore what \emph{effective} energy-momentum tensor would yield this particular solution in general relativity to provide for an analogous view on it. In Section~\ref{sec:BHProps} we study properties of the black hole solutions and how they differ from Schwarzschild geometry. Classical observables derived from the motion of particles are discussed in Section~\ref{ssec:ppmotion}, the horizon of the black hole in Section~\ref{ssec:BHhor}, the surface gravity and black hole temperature in Section~\ref{ssec:BHtemp}, and, finally, sparsity in Section~\ref{ssec:spars}. We end the paper with concluding remarks in Section~\ref{sec:conc}.

The index conventions used in this article are that Greek indices label spacetime coordinate indices and Latin indices label Lorentz frame components. The metric has the signature $(+,-,-,-)$. We use geometric units in which $c=\hbar=k_{\text{B}}=1$, Newton's constant $G$ is retained.

\section{\texorpdfstring{Black holes in weak $f(\mathbb{T})$ gravity}{Black holes in weak f(T) gravity}}\label{sec:fTBH}
After briefly recalling the notions of covariant $f(\mathbb{T})$ gravity, we display again the general static spherically symmetric vacuum solution of weak $f(\mathbb{T})$ gravity, which has been found in earlier studies. We discuss why these previously discussed solutions are not the most general black hole solutions for this modified theory of gravity. Additionally, they cannot necessarily be interpreted as perturbations of a Schwarzschild black hole.

We start the construction of such solutions from demanding a finite determinant of the metric at the horizon, and find the general black hole solutions of weak $f(\mathbb{T})$ gravity.

\subsection{\texorpdfstring{Covariant $f(\mathbb{T})$ gravity}{Covariant f(T) gravity}}\label{ssec:covfT}
In covariant teleparallel gravity \cite{TPGravBook,Golovnev:2017dox,Krssak:2015oua,Hohmann:2017duq}, the fundamental fields encoding the gravitational dynamics are a tetrad $\theta^a$, and a flat, metric compatible spin connection $\omega^a{}_{b}$ with torsion $T^a$. The tetrad fields satisfy
\begin{align}
	\theta^a = h^a{}_{\mu}\dif x^\mu, \quad e_a = h_a{}^\mu \partial_\mu,\quad \theta^a(e_b) = \delta^a_b \quad \Rightarrow \quad g_{\mu\nu} = \eta_{ab}h^a{}_\mu h^b{}_\nu \,,
\end{align}
where $\eta_{ab}$ is the Minkowski metric.  The spin connection is generated by local Lorentz matrices~$\Lambda^a{}_b$
\begin{align}
    \omega^a{}_{b\mu} = \Lambda^a{}_c \partial_\mu (\Lambda^{-1})^c{}_b\,,
\end{align}
and its torsion is given by
\begin{align}
    T^a{}_{\mu\nu} = 2 \left( \partial_{[\mu}h^a{}_{\nu]} + \omega^a{}_{b[\mu}h^b{}_{\nu]}\right)\,.
\end{align}
To transform an index from a Latin Lorentz index to a Greek coordinate index, or vice versa, contractions with the components of a tetrad $h^a{}_\mu$, respectively inverse tetrad $h_a{}^\mu$ are applied.

The Lorentz matrices are pure gauge fields and, without loss of generality, it is possible to work in the so called Weitzenböck gauge, in which one absorbs the Lorentz matrices in the tetrad. As a consequence one can globally work with zero spin connection in this gauge, see, for example, \cite[Eq. (4)]{Bahamonde:2020snl} for a detailed derivation. Throughout the rest of this article we will work in Weitzenböck gauge in which the torsion becomes $T^a{}_{\mu\nu} = 2  \partial_{[\mu}h^a{}_{\nu]}$.

Teleparallel theories of gravity are defined in terms of an action, 
\begin{align}
    S[h] = \int \dif^4x |h| \mathcal{L}_{\text{G}} + \mathcal{L}_{\text{M}}(h,\Psi)
\end{align}
whose gravitational Lagrangian $\mathcal{L}_{\text{G}}$ is constructed from scalars built in terms of the torsion tensor. The matter Lagrangian $\mathcal{L}_{\text{M}}(g,\Psi)$ is assumed to depend on the tetrad only through the metric coupling the matter fields $\Psi$ to gravity.

The most fundamental, parity even building blocks for the gravitational action are the three quadratic scalars $T_1 = T^\sigma{}_{\mu\nu} T_{\sigma}{}^{\mu\nu}$, $T_2 = T^\sigma{}_{\mu\nu}T^{\nu\mu}{}_\sigma$ and $T_{3} = T^{\sigma\mu}{}_{\sigma}T^{\rho}{}_{\rho\mu}$. From these one can construct the torsion scalar
\begin{align}
    \mathbb{T} = T^{a}{}_{\mu\nu} S_{a}{}^{\mu\nu}= \frac{1}{4} T_1 + \frac{1}{2} T_2 + T_3\,,
\end{align}
where $S_a{}^{\mu\nu} = \frac{1}{2}(K^{\mu\nu}{}_a - h_a{}^\mu T_\lambda{}^{\lambda\nu} + h_a{}^\nu T_\lambda{}^{\lambda\mu})$ is the so-called superpotential, in turn given in terms of the contortion tensor $K^{\mu\nu}{}_a = \frac{1}{2}(T^{\nu\mu}{}_a + T_a{}^{\mu\nu} - T^{\mu\nu}{}_a)$.

Setting $\mathcal{L}_{\text{G}} = \frac{1}{2\mathfrak{K}}\mathbb{T}$, with $\mathfrak{K} = 8\pi G$, defines the teleparallel equivalent of general relativity (TEGR), the teleparallel theory of gravity which is dynamically equivalent to general relativity. Setting instead $\mathcal{L}_{\text{G}} = \frac{1}{2\mathfrak{K}}f(\mathbb{T})$ defines $f(\mathbb{T})$-gravity. Variation of the action with respect to the tetrad components yields the field equations
\begin{align}\label{eq:fT}
	\frac{1}{4}f(\mathbb{T}) h_a{}^\mu +  f_\mathbb{T}\ \left( T^b{}_{\nu a} S_b{}^{\mu \nu } + \frac{1}{h}\partial_{\nu}(h S_a{}^{\mu \nu }) \right) +  f_{\mathbb{T}\mathbb{T}}\  S_a{}^{\mu\nu} \partial_\nu \mathbb{T} &= \frac{1}{2}\mathfrak{K} \Theta_a{}^\mu\,.
\end{align}
After contraction with tetrad components and lowering an index with the spacetime metric this can be cast into the form $H_{\sigma\rho} = \frac{1}{2}\mathfrak{K} \Theta_{\sigma\rho}$, and can then be separated into symmetric and anti symmetric parts
\begin{align}\label{eq:fT2}
	 H_{(\sigma\rho)} = \frac{1}{2}\mathfrak{K} \Theta_{(\sigma\rho)}, \quad  H_{[\sigma\rho]} =0\,.
\end{align}

In the following we consider the theory going by the name of \enquote{weak $f(\mathbb{T})$-gravity}. It is defined by specifying the free function $f$ to be
\begin{eqnarray}\label{eq:deff}
f(\mathbb{T})=\mathbb{T}+\frac{1}{2}\alpha \epsilon\, \mathbb{T}^2\,,
\end{eqnarray}
where $\alpha$ is a coupling parameter, and $\epsilon$ a perturbation parameter for bookkeeping purposes. Later, a dimensionless parameter $\beta$ also containing the original, general-relativistic Schwarzschild radius $r_{\text{s}}$ will be used instead of $\alpha$.

\subsection{\texorpdfstring{Static spherically symmetric black holes in weak f($\mathbb{T}$) gravity}{Static spherically symmetric black holes in weak f(T) gravity}}\label{ssec:stspBHs}
In weak $f(\mathbb{T}$)-gravity, a general, static, spherically symmetric vacuum solution family to the field equations~\eqref{eq:fT} has been found in \cite{DeBenedictis:2016aze,Bahamonde:2019zea}. In these articles, special asymptotically flat solutions have been further studied. We now recall the general, static, spherically symmetric vacuum solution and argue that the boundary conditions chosen in the previous studies do not necessarily lead to black hole solutions. Afterwards we identify all black hole solutions of weak $\mathbb{T}$-gravity.

\subsubsection{The general static spherically symmetric solution}\label{ssec:ssym}
Employing the standard static spherically symmetric tetrad in spherical coordinates $(t,r,\theta,\phi)\in \mathbb{R}\times (0,\infty) \times (0,\pi) \times (0,2\pi)$, which is compatible with vanishing spin connection, \cite{Bohmer:2011si,Hohmann:2019nat},
\begin{equation}
h^a{}_{\nu}=\left(
\begin{array}{cccc}
\sqrt{A} & 0 & 0 & 0 \\
0 & \sqrt{B} \cos (\phi ) \sin (\theta ) & r \cos (\phi ) \cos (\theta )  & -r \sin (\phi ) \sin (\theta )  \\
0 & \sqrt{B} \sin (\phi ) \sin (\theta )  & r \sin (\phi ) \cos (\theta )  & r \cos (\phi ) \sin (\theta ) \\
0 & \sqrt{B} \cos (\theta ) & -r \sin (\theta ) & 0 \\
\end{array}
\right)\label{eq:tetrad}
\end{equation}
immediately solves the antisymmetric part of the field equations~\eqref{eq:fT2}. The metric induced by this tetrad is the standard static and spherically symmetric one
\begin{align}\label{eq:sphmet}
    g = A \dif t^2 - B \dif r^2 - r^2 (\dif \theta^2 + \sin^2\theta \dif\phi^2)\,.
\end{align}

The symmetric vacuum field equations~\eqref{eq:fT2} for weak $f(\mathbb{T})$-gravity are solved as a first order perturbation around Schwarzschild geometry by
\begin{alignat}{1}
    A(r)= \mu^2 +&\epsilon\left(- \frac{C_1 (1-\mu^2)}{r_{\text{s}}}+C_2 \right.\nonumber\\
        &~~\left.- \beta \frac{(51-93 \mu^2 -128 \mu^3 + 45 \mu^4 - 3 \mu^6 - 12 (1-3 \mu^2) \ln(\mu))}{6}\right)\,,\label{afinX}\\
    B(r)= \frac{1}{\mu^2}+ &\epsilon \frac{(1-\mu^2)}{\mu^4} \left(\left[\frac{C_1}{r_{\text{s}}} - C_2\right]\right. \nonumber\\
        &\left.+ \beta \frac{(63 -24 \mu + 12 \mu^2 +64 \mu^3 - 75 \mu^4 + 24 \mu^5 - 12 \ln(\mu))}{6}\right)\,.\label{bfinX}
\end{alignat}
where $\mu=(1-r_{\text{s}}/r)^{1/2}$,  $r_{\text{s}}$ is the Schwarzschild radius, $\beta = \frac{\alpha}{r_{\text{s}}^2}$, and the integration constants $C_1$ and $C_2$ need to be determined by suitable boundary conditions. The details for the derivation of the solution can be found in \cite{DeBenedictis:2016aze,Bahamonde:2019zea}. 

Let us shortly summarize their results: In these latter articles, an expansion in $\frac{1}{r}$  was used to determine the constants of integration to be $C_1 = -32 r_{\text{s}} \beta$ and $C_2 = - \frac{64}{3} \beta$ such that in the asymptotic region $r\to\infty$ the first non-vanishing order of the metric coefficients are of order $\frac{1}{r^2}$. It is worth stressing that this fixing of the integration constants leaves no freedom to ensure that important necessary conditions perturbative black hole solutions have to satisfy are fulfilled near the central mass. The resulting metric coefficients are
\begin{align}
    A(r(\mu)) &= \mu^2 +  \frac{\beta}{6} (13 - 99 \mu^2 + 128 \mu^3 - 45 \mu^4 + 3 \mu^6 + 12(1-3\mu^2) \ln(\mu))\,,\label{afinX2Old}\,\\
    B(r(\mu)) &= \frac{1}{\mu^2} - \beta (1-\mu^2) \frac{1 + 24 \mu - 12 \mu^2 - 64 \mu^3 + 75 \mu^4 - 24 \mu^5 + 12 \ln(\mu)}{6 \mu^4}\,.\label{bfinX2Old}
\end{align}

In order to interpret a spherically symmetric metric as a black hole metric several necessary conditions need to be satisfied. One important condition is that the determinant of the metric~\eqref{eq:sphmet} 
\begin{align}
    \det g = - A B r^2 \sin^2\theta\,.
\end{align}
must be non-degenerate at the black hole horizon at $r=r_{\text{h}}$.\footnote{As we work in spherical coordinates, this determinant also mirrors the fact that one needs more than one coordinate patch to cover the sphere, concretely, the singularities at the poles.} This implies that $A(r_{\text{h}}) B(r_{\text{h}}) = \zeta >0$, for some fixed constant $\zeta$, must hold at the putative horizon where $A(r_{\text{h}})=0$. For our perturbative ansatz
\begin{align}\label{eq:pertans}
    A(r) = a_0(r) + \beta a_1(r),\quad B(r) = b_0(r) + \beta b_1(r)\,,
\end{align}
the vanishing of the $tt$ component of the metric implies (to first order in $\beta$ and $\beta\neq0$)
\begin{align}
    A(r_{\text{h}}) &= 0 \Rightarrow a_0(r_{\text{h}}) = - \beta a_1(r_{\text{h}})
\end{align}
and, using also a first order expansion $\zeta = 1 + \beta \zeta_1$,
\begin{align}
    A(r_{\text{h}}) B(r_{\text{h}}) &= a_0(r_{\text{h}}) b_0(r_{\text{h}}) + \beta (a_0(r_{\text{h}}) b_1(r_{\text{h}}) + b_0(r_{\text{h}}) a_1(r_{\text{h}})) = 1 + \beta \zeta_1 \\
    \mathclap{\Rightarrow\qquad\qquad\qquad\qquad} b_0(r_{\text{h}}) &= a_0(r_{\text{h}})^{-1},\quad b_1(r_{\text{h}}) = \frac{\zeta_1}{a_0(r_{\text{h}})} - \frac{a_1(r_{\text{h}})}{a_0^2(r_{\text{h}})}\,.
\end{align}
This finally yields $B(r_{\text{h}}) = \frac{2 + \beta \zeta_1}{a_0(r_{\text{h}})} +\mathcal{O}\left(\beta^2\right)$. This represents a necessary condition for a well-defined, perturbative black hole solution in weak $f(\mathbb{T})$-gravity.

From this analysis we see that the  metric coefficients~\eqref{afinX2Old} and~\eqref{bfinX2Old} superficially seem to satisfy this perturbative condition to be a black hole, for which the parameter $\zeta_1$ is non-zero and given by
\begin{align}
    \zeta_1 = -\frac{\mu_{\text{h}}-1}{3 \mu_{\text{h}}^2}(6 - 6 \mu_{\text{h}} - 49 \mu_{\text{h}}^2 + 59 \mu_{\text{h}}^3 - 7 \mu_{\text{h}}^4 - 27 \mu_{\text{h}}^5 + 12 \mu_{\text{h}}^6) - 4 \ln(\mu_{\text{h}})\,.
\end{align}
It is positive for $0<\mu_{\text{h}}<1$. However, the product $\beta \zeta_1(\mu_{\text{h}})$ is not necessarily small against $1$ for $\mu_{\text{h}}(\beta)$, which can be determined numerically by solving~\eqref{afinX2Old}, \textit{i.e.}, $A(\mu_{\text{h}})=0$ for $\mu_{\text{h}}$ for fixed $\beta \neq 0$. This shows that choosing both integration constants $C_1$ and $C_2$ to fix the fall-off properties of the metric coefficients at $r\to\infty$ does not necessarily lead to a perturbative treatment of the Schwarzschild black hole spacetime near the horizon. A perturbative analysis has to leave open whether or not these particular choices represent black hole solutions or not.

\subsubsection{The black hole solution}\label{ssec:BHs}
Let us therefore look for perturbative solutions that are more amenable to an interpretation as a black hole. In order to do this for weak $f(\mathbb{T})$-gravity we determine the integration constant $C_2$ such that $A(r_{\text{h}}) B(r_{\text{h}})=1 + \beta \zeta_1 $ is satisfied. Note that unlike before this involves not only a condition at infinity (the requirement to be comparable to Schwarzschild), but also one at the \emph{putative} horizon. This condition yields
\begin{align}\label{eq:c2}
	C_2 = \beta\frac{\left(-6+12 \mu_{\text{h}} - 21 \mu_{\text{h}}^2 - 108 \mu_{\text{h}}^3 + 66 \mu_{\text{h}}^4 + 20 \mu_{\text{h}}^5 - 39 \mu_{\text{h}}^6 + 12 \mu_{\text{h}}^7 + 12 \mu_{\text{h}}^2 \ln(\mu_{\text{h}}) + 3 \mu_h^2 \zeta_1 \right)}{3 \mu_{\text{h}}^2} \,,
\end{align}
where $\mu_{\text{h}}=(1-r_{\text{s}}/r_{\text{h}})^{1/2}$ is $\mu$ evaluated at the horizon radius $r_{\text{h}}$, and $\zeta_1$ is the above defined, first-order in $\beta$ contribution to the determinant of the metric at the horizon. The remaining integration constant $C_1$ in equations~\eqref{afinX} and~\eqref{bfinX} is determined by demanding appropriate fall-off behaviour of the metric components for large $r$ for regaining the Schwarzschild solution in the limit $\beta\to 0$. The resulting expansion of the metric coefficients in $\frac{1}{r}$ yields
\begin{align}
    A(r) &= 1-\frac{r_{\text{s}}}{r} + \epsilon \left( C_2 + \frac{64}{3} \beta - \frac{( C_1 + 32 r_{\text{s}} \beta)}{r} \right) + \mathcal{O}\left(\frac{1}{r^2}\right)\\
    B(r) &= \left(1-\frac{r_{\text{s}}}{r}\right)^{-1} + \epsilon \frac{1}{r}\left( C_1 - C_2 r_{\text{s}} + \frac{32 r_{\text{s}}}{3} \right) + \mathcal{O}\left(\frac{1}{r^2}\right)\,.
\end{align}
Thus, since $C_2$ is already determined from the finiteness of the determinant at the horizon, the best choice to minimize the teleparallel influence in the asymptotic region is to demand that $B(r) = 1+\frac{r_{\text{s}}}{r} + \mathcal{O}\left(\frac{1}{r^2}\right)$. This in turn implies
\begin{align}
    C_1 = C_2 r_{\text{s}} - \frac{32}{3}\beta\,.
\end{align}
This concludes our search for appropriate integration constants. Our final choice then yields a perturbative, weak $f(\mathbb{T})$-gravity, 2-parameter family of black holes defined by 
\begin{align}
    A(r(\mu)) &= \mu^2 +  \frac{\beta}{6} (13 + 29 \mu^2 + 128 \mu^3 - 45 \mu^4 + 3 \mu^6 + 12(1-3\mu^2) \ln(\mu)) + C_2 \mu^2 \,,\label{afinX2BH}\,\\
    B(r(\mu)) &= \frac{1}{\mu^2} + \beta \frac{\mu^2-1}{6 \mu^4} \left( 1 + 24 \mu + 12 \mu^2 + 64 \mu^3 - 75 \mu^4 + 24 \mu^5 - 12 \ln(\mu)\right) \,.\label{bfinX2BH}
\end{align}
In the $M\to0$ limit\footnote{This corresponds to $r_{\text{s}}\to0$, and hence to $\mu\to 1$ and $\mu_{\text{h}}\to1$.}, we obtain $B(r) = 1$ and $ A(r) = 1 + \zeta_1$, \textit{i.e.}, one obtains Minkowski spacetime by a simple redefinition of time coordinate $t\to \sqrt{1+\zeta_1}t$. As a constant rescaling, this does not affect the the spin connection, \textit{i.e.}, the rescaled tetrad is still in Weitzenböck gauge.

The asymptotic behaviour of this solution is as follows. For large $r$ the dominating terms are $B(r) = 1 - \frac{r_{\text{s}}}{r} + \mathcal{O}\left(\frac{1}{r^2}\right)$ by construction, and 
\begin{align}
    A(r) = \left(1 - \frac{r_{\text{s}}}{r}\right) \left( 1 +   C_2 + \beta \frac{64}{3} \right) + \mathcal{O}\left(\frac{1}{r^2}\right).
\end{align}
Hence the teleparallel corrections to general relativity for the weak teleparallel black hole seem to become relevant in the $r\to \infty$ region. However---and as mentioned above---, a simple constant rescaling (thus not affecting the spin connection) of the time coordinate $t\to \sqrt{1+C_2 + \beta \frac{64}{3}} t$ makes the metric for large $r$ identical to the Schwarzschild metric in leading order.

Thus we found the most general family of static asymptotically flat black hole solutions of weak $f(\mathbb{T})$-gravity, which are parametrized by the zeroth order parameter $r_{\text{s}}$, by the perturbation parameter $\beta$ and the horizon parameter $\zeta_1$. In case one determines the constant of integration $C_2$ by other means the value of $\zeta_1$ is fixed by equation~\eqref{eq:c2}. Choosing $\zeta_1=0$ ensures that the determinant of the metric at the horizon of the weak $f(\mathbb{T})$-black hole will have the same value as the Schwarzschild black hole at the Schwarzschild horizon. It is important to note that while our family of \enquote{black hole solutions} works well for $r>r_{\text{h}}$ perturbatively, this perturbative solution has limitations at the horizon itself. This is still a strong improvement over the earlier static and spherically symmetric solutions, where the perturbative ansatz breaks down much earlier when looking at the determinant of the metric.

\subsection{The general relativistic perspective---energy conditions}\label{sec:gr}

Any modified theory of gravity can, as long as a metric is its outcome, be re-examined from the viewpoint of general relativity. Obviously, what was a vacuum spacetime in the modified theory is usually not a general relativistic vacuum spacetime. This will also be true for our case of weak $f(\mathbb{T})$-gravity. Nonetheless, it can often help our intuition to look at a given metric through the lens of GR again. This study through analogy is well-known in the field of modified theories of gravity (see \cite{Alvarenga:2012bt,Capozziello2015GenEnergyCondModGrav} and references therein).

A key observation to do this is that one can use \eqref{eq:fT} to rewrite the symmetric field equations \eqref{eq:fT2} with help of the Einstein tensor $G_{\mu\nu}$ as
\begin{align}
\begin{split}
    G_{\mu\nu} 
    &= R_{\mu\nu} - \frac{1}{2}g_{\mu\nu}R, \\
    &= - \beta r_s \left( \frac{1}{4} \mathbb{T}^2 g_{\mu\nu} + 2 \left(-T_{b\sigma(\mu}S^{b\sigma}{}_{\nu)} + \frac{1}{h}h^a{}_{(\mu} g_{\nu)\sigma} \partial_\lambda(h S_a{}^{\sigma\lambda}) \right) \mathbb{T} + 2 S_{(\mu\nu)}{}^{\sigma} \partial_\sigma \mathbb{T} \right), \\ 
    &= T_{\mu\nu}^{\text{eff}}(T^a{}_{\mu\nu},T^a{}_{\mu\nu,\sigma}),
\end{split}
\end{align}
where we used $\Theta_{\mu\nu}=0$ and $f(\mathbb{T}) = \mathbb{T} + \frac{1}{2}\beta r_s\mathbb{T}^2$ as well as the fact that for $f(\mathbb{T}) = \mathbb{T}$, \emph{i.e.}, TEGR, the field equations are identical to the ones of general relativity \cite{TPGravBook}.  

Thus, a metric derived from the vacuum equations of a modified theory of gravity can be interpreted in GR as a particular matter source in the form of an \emph{effective} energy-momentum tensor $T_{\mu\nu}^{\text{eff}}$.\footnote{Whether or not this matter source would have a good physical interpretation is outside of the scope of the current article.} The energy conditions available in GR provide a valuable toolbox for a first check if and how a given solution of a modified theory of gravity could be plagued by pathologies.

In case the solution yields an effective energy momentum tensor which satisfies the energy conditions, theorems in GR exclude many pathologies \cite{HawEll,VisserWormholes,CurielPrimerEnergyConditions,Lobo2017WormholesETC}. This will prove particularly useful for solutions of modified gravity that are more involved than the one presented here, nevertheless as pedagogic proof of principle, we now present this line of reasoning.

In our case, for the metric given by equations~\eqref{eq:sphmet},~\eqref{afinX2BH}, and~\eqref{bfinX2BH}, the Einstein tensor in the canonical diagonal tetrad basis of the the metric~\eqref{eq:sphmet}) becomes
\begin{subequations}\label{eq:Gtetrad}
    \begin{align}
        G_{\hat{t}\hat{t}} &= \beta \frac{(\mu-1)^7(\mu+1)^4(1 + 5\mu + 10\mu^2)}{r_{\text{s}}^2 \mu^2}, \\
        G_{\hat{r}\hat{r}} &= \beta \frac{(\mu - 1)^8 (\mu+1)^4}{r_{\text{s}}^2 \mu^2},\\
        G_{\hat{\theta}\hat{\theta}} &= -\beta \frac{(\mu-1)^8(\mu+1)^2(1+4\mu+5\mu^2)}{2r_{\text{s}}\mu^3},\\
        G_{\hat{\phi}\hat{\phi}} &= G_{\hat{\theta}\hat{\theta}}.
    \end{align}
\end{subequations}
Obviously, this is diagonal---in other words, this is a type~I energy-momentum tensor according to the Hawking--Ellis (a.k.a. Segré--Plebański) classification \cite{HawEll}. Even without explicitly calculating the Einstein tensor, a general result for static and spherically symmetric spacetimes \cite{MartinMorunoVisser2021HawkingEllisClassification} would tell us this in advance. With equations~\eqref{eq:Gtetrad} at hand, it is straightforward to check the point-wise energy conditions \cite{CurielPrimerEnergyConditions,MartinMoruniVisser2017EnergyConditions}. From these one can immediately read off the energy density and pressures entering the energy conditions (in what calls Curiels their \enquote{effective form}).\footnote{For the precise formulation, we refer the reader to \cite{CurielPrimerEnergyConditions}. Note the different sign convention for the metric enters the direction of the inequalities.} It quickly follows that the null energy condition, weak energy condition, strong energy condition, and dominant energy condition are all fulfilled. This is not \emph{too} surprising given how relatively benign our metric is.

\section{Classical and semi-classical properties}\label{sec:BHProps}
Having found static, spherically symmetric black hole solutions of weak $f(\mathbb{T})$-gravity, we now analyse properties of these solutions for $1\gg\beta \zeta_1$ to compare these to the ones given in the literature. This will lead to a new point of view regarding the observational constraints on teleparallel gravity. This is done best by looking at the behaviour of classical point particle trajectories, such as the photon sphere around the black hole, the perihelion shift, the Shapiro delay, and the scattering of light.

Moreover we study properties of the black hole which are connected to its horizon, such as the surface gravity, and---as a semi-classical extension---the sparsity, which characterize aspects of Hawking radiation.

\subsection{Particle propagation effects: Photon sphere, perihelion shift, Shapiro delay and light deflection}\label{ssec:ppmotion}
The motion of point particles in spherical symmetric spacetimes---their geodesic equation---is derived from the Lagrangian,
\begin{align}
    \mathcal{L} = \frac{1}{2}g_{\mu\nu}\dot x^\mu \dot x^\nu = A \dot t^2 - B \dot r^2 - r^2 (\dot \theta^2 + \sin^2\theta \dot\phi^2)\,.
\end{align}
There exist two constants of motion, the energy $E= \frac{\partial \mathcal{L}}{\partial \dot t}$ and the angular momentum $L= \frac{\partial \mathcal{L}}{\partial \dot \phi}$, and, thanks to spherical symmetry, without loss of generality we can restrict the analysis to the equatorial plane $\theta = \frac{\pi}{2}$. This results in the fact that the sole remaining equation of motion to solve is
\begin{align}\label{eq:geod}
    \frac{1}{2}\dot r^2 + V(r) = 0\,,
\end{align}
where the effective potential for the perturbative metric coefficients $A(r) = (1-\frac{r_{\text{s}}}{r}) + \beta a(r)$ and $B(r) = (1-\frac{r_{\text{s}}}{r})^{-1} + \beta b(r)$ is, to first order in $\beta$, given by
\begin{align}
   	V(r) =& - \frac{1}{2} E^2 + \frac{1}{2} \left(1-\frac{r_{\text{s}}}{r}\right) \left( \frac{L^2}{r^2} + \sigma \right) \nonumber\\
	&+ \frac{\beta}{2} \left[ E^2 \left( \frac{a(r)}{1-\frac{r_{\text{s}}}{r}} + b(r)\left(1-\frac{r_{\text{s}}}{r}\right)\right) - b(r)\left( \sigma + \frac{L^2}{r^2} \right)\left(1-\frac{r_{\text{s}}}{r}\right)^2\right]\,.
\end{align}
Here, $\sigma=0$ for massless and $\sigma=1$ massive particles. The functions $a(r)$ and $b(r)$ for the weak $f(\mathbb{T})$ black hole can be extracted from~\eqref{afinX2BH} and~\eqref{bfinX2BH}.

With this in place, it is possible to calculate various classical observables for our black hole spacetime. Details on the derivation of these can for example be found in the textbooks \cite{Weinberg,FrolovZelnikov2011}.

\begin{itemize}
    \item {\textbf{Photon sphere:}} The photon sphere---a characterizing feature of black holes \cite{Virbhadra:1999nm,Claudel:2000yi}---is derived from the geodesic equations~\eqref{eq:geod} by searching for orbits with $\dot r=0$ and $\sigma=0$. For this, we solve $V(r)=0$ and $V(r)'=0$ to first order in $\beta$. 
    
    In our case the photon sphere then lies at
    \begin{align}
        r_{\text{ph}} = \left(\frac{3}{2} + \beta \frac{(27 \ln(3) + 32 \sqrt{3} - 80)}{18} \right) r_{\text{s}} \approx \left( 1.5 + \beta 0.282675 \right) r_{\text{s}}\,.
    \end{align}
    This result is identical to the one found in~\cite{Bahamonde:2019zea}.
    
    \item {\textbf{Perihelion shift:}} While an elliptic orbit in the Newtonian two-body problem would experience the perihelion always at the same angle, deviations from the two-body problem---either by adding more bodies or, as here, using different dynamics---forces this perihelion to move from orbit to orbit. For sufficiently small eccentricity, this is encoded in the quantity $\Delta \phi$ of an orbit $r(\phi)=r_{\text{c}}+r_{\phi}(\phi)$ which is a perturbation of an orbit with constant radius $r_{\text{c}}$. It is derived from the effective potential as, see for example \cite{Bahamonde:2019zea} for a derivation, 
    \begin{align}
        \Delta \phi =2\pi \left(\frac{h}{r_{\text{c}}^2\sqrt{V''(r_{\text{c}})}}-1\right)\,.
    \end{align}
    To first order in $\beta$, we find
    \begin{align}
        \Delta \phi = 6 \pi  q + 27 \pi  q^2  + 135 \pi q^3 + \frac{2835 \pi}{4} q^4 + \beta 32 \pi q^4 +\mathcal{O}(q^5)\,,
    \end{align}
    where $q = \frac{r_{\text{s}}}{2 r_{\text{c}}}$ is an expansion parameter, assumed to be small, \textit{i.e.}, $r_{\text{c}} \gg r_{\text{s}}$. The influence of the teleparallel perturbation of GR on the perihelion shift in the black hole spacetimes constructed in this paper is much weaker than the result found in~\cite{Bahamonde:2019zea,DeBenedictis:2016aze}, where a correction already appeared at order~$q^2$.
    
    This finding thus leads us to weaker constraints on the teleparallel coupling $\alpha = \beta r_{\text{s}}^2$ from observations of the perihelion shift, for example, from the orbits of stars around the black hole in the center of our galaxy.

    \item {\textbf{Shapiro time delay:}} The Shapiro time delay is the time delay experienced by a radar signal between an emitter at $r_{\text{e}}$ and a mirror at $r_{\text{m}}$ due to the presence of a gravitational mass \cite{Shapiro}. The time which passes until a light ray has travelled from an emitter $r=r_{\text{e}}$ to a point of closest encounter to the gravitational mass at $r_0$, respectively, from $r_0$ to a mirror at $r_{\text{m}}$, is given by
    \begin{align}
        t(r_X,r_0) 
        &= \int_{r_0}^{r_X} \frac{E}{\sqrt{-2 V(r)}} \dif r, \qquad  X=\text{e,m} \\
        &= t(r_X,r_0)_{\text{GR}}\ \left(1 - \frac{\zeta_1}{2} \beta\right) + \mathcal{O}\left( \left(\frac{r_{\text{s}}}{r_X}\right)^5\right)\,.
    \end{align}
    The last equality holds when the integrand is expanded in powers of the small parameter $\frac{r_{\text{s}}}{r}$, \textit{i.e.}, assuming $r_0>r_{\text{s}}$. The lowest orders can be integrated explicitly, but the higher orders do not allow for a (known) closed formula for the integral.
    
    We further assume that changes in the relative distances between emitter, mirror and mass can be neglected during this propagation. The total travel time for a return trip of the light signal is then $\Delta t = 2 (t(r_{\text{e}},r_0) + t(r_{\text{m}},r_0) )$. The Shapiro delay is given by
    \begin{align}
        \Delta t_{\text{Shapiro}} 
        &= \Delta t - 2 \left( \sqrt{r_{\text{e}}^2 - r_0^2} + \sqrt{r_{\text{m}}^2 - r_0^2} \right))\left(1 - \frac{\zeta_1}{2} \beta\right) \\
        &= \Delta t_{\text{Shapiro,GR}}\ \left(1 - \frac{\zeta_1}{2} \beta\right)+ \mathcal{O}\left( \left(\frac{r_{\text{s}}}{r_X}\right)^5\right)\,,
    \end{align}
    where $\sqrt{r_X^2 - r_0^2}\left(1 - \frac{\zeta_1}{2} \beta\right)$ is the travel time of the light ray in the absence of the gravitating mass. Here, we see explicitly the influence of the horizon parameter. It changes the units of time measurements, however this can easily be absorbed by a coordinate rescaling of the time coordinate, as we discussed already below at the end of Section~\ref{ssec:BHs}. A non-trivial contribution to the Shapiro delay only emerges at fifth order in the small parameter $\frac{r_x}{r_{\text{s}}}$, for which it is, however, not possible to analytically evaluate the above integral.
    
    \item {\textbf{Light deflection:}} Another central quantity to investigate in this context is the deviation angle $\Delta \phi$ between a null geodesic in the presence and the absence of a central gravitating mass. The point of closest encounter to the central object is again called $r_0$. Then the deflection angle is given by
    \begin{align}
        \Delta \phi = 2 \int_{r_0}^\infty \left(\frac{L}{r^2 \sqrt{-2V(r)}} \dif r \right) - \pi = \Delta_{\phi\text{,GR}} + \beta \frac{4}{15}\frac{r_{\text{s}}^5}{r_0^5}\,.
    \end{align}
    where in the last equality an expansion for $r_0 \gg r_{\text{s}}$ is employed in order to evaluate the integral. The first non-vanishing correction to general relativity appears again only at fifth order in the small parameter $\frac{r_{\text{s}}}{r_0}$.

    \item {\textbf{Minimal photon impact parameter:}} The impact parameter $b$ is another closely related and only mildly different way to look at these matters of light deflection. It adds however, a simple way to characterize the photon capture cross-section of the black hole for an observer at infinity. Physically and more concretely, the \emph{minimal} photon impact parameter $b_{\text{min}}$ characterizes the closest encounter with the central mass a photon can have when it is scattered by this mass, and can still be received by a distant observer. It is defined as follows. Define $\ell = \frac{L}{E r_{\text{s}}}$ and demand that the effective potential $V$ as a function $V=V(r,E, L(E, \ell))$ vanishes, which determines the ratio between $E$ and $L$ as a function of $r$, \textit{i.e.}, $\ell=\ell(r)$. This quantity possesses a minimum at $r=r_{\text{ph}}$ which identifies the minimal possible impact parameter for scattering as
    \begin{align}
        b_{\text{min}} = \sqrt{\ell(r_{\text{ph}})} = \frac{3 \sqrt{3}}{2} - \frac{1}{4} \beta \left(64 + \frac{80}{\sqrt{3}} + 3 \sqrt{3} \frac{C_2}{\beta} \right)\,.
    \end{align}
    Every geodesic with smaller impact parameter $b$ will be gravitationally captured. We introduce $b_{\text{min}}$ here, since it is needed for the discussion of the black hole sparsity in Section~\ref{ssec:spars}. Unlike in the previous cases, no additional small parameter was introduced.
    
    The minimal photon impact parameter $b_{\text{min}}$ is the crucial quantity to describe the shadow of these black holes. The shadow corresponds exactly to the capture cross-section. Stricly speaking, the phrase \enquote{silhouette} would be a more apt description of this capture cross-section, but the phrase \enquote{shadow} is the established terminology, and we will abide by it.
    
\end{itemize}

We see that the photon sphere is affected strongest by the teleparallel perturbation, the effect on the perihelion shift appears at orders $\left(\frac{r_{\text{s}}}{r}\right)^4$ for large $r$ while an effect on the Shapiro delay and on the perihelion shift emerges at powers $\left(\frac{r_{\text{s}}}{r}\right)^5$ for large r. The modification are significantly smaller than the corrections induced by the asymptotically flat solutions reported in \cite{DeBenedictis:2016aze,Bahamonde:2019zea}.

Our findings nicely quantify the statement \enquote{the outer region, in turn, is indistinguishable from the Schwarzschild spacetime provided the Schwarzschild mass satisfies $\frac{r_{\text{s}}^2}{4} \gg \frac{1}{|\lambda|}$}, made in the context of Black holes in Born--Infeld gravity \cite{Boehmer:2020hkn}, where their parameter $\frac{1}{\lambda}$ is our parameter $\alpha$ as employed in~\eqref{eq:deff}.

\subsection{The event horizon}\label{ssec:BHhor}
The putative horizon of the weak $f(\mathbb{T})$-gravity black hole we are investigating is given by the solution of the equation, see~\eqref{afinX2BH},
\begin{align}
    A(r(\mu_{\text{h}})) = \mu_{\text{h}}^2 +  \frac{\beta}{6} [13 - 99 \mu_{\text{h}}^2 + 128 \mu_{\text{h}}^3 - 45 \mu_{\text{h}}^4 + 3 \mu_{\text{h}}^6 + 12(1-3\mu_{\text{h}}^2) \ln(\mu_{\text{h}})] + C_2 \mu_{\text{h}}^2 = 0\,.
\end{align}
Sadly it is neither possible to solve this equation analytically, nor in a simple linear perturbative series. Due to the presence of the logarithmic terms, the Newton--Raphson method cannot yield a good linear approximation for the position of the horizon in $\beta$. We therefore decided to calculate $r_{\text{h}}$ numerically for fixed values of $\beta$ and $\zeta_1$ at much higher precision. In subsequent quantities depending on $r_{\text{h}}$, we checked that the Schwarzschild limit $\beta\to 0$ comes out correctly, and first order terms in $\beta$ have smaller absolute values than the zeroth order, Schwarzschild contribution. Put differently, when an evaluation at $r_{\text{h}}$ was required, as described above, we treated that term as a function of $r$, approximated it, and \emph{then} evaluated it at the putative horizon. For the rest of this article, we will choose $\zeta_1=0$ in all numerical calculations. 

\begin{figure}[h!]
    \centering
    \includegraphics[width=.66\textwidth]{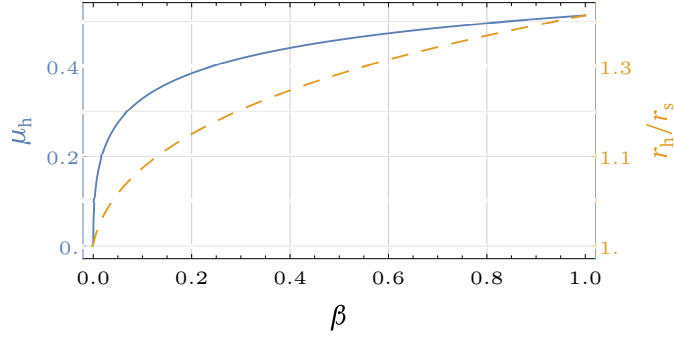}
    \caption{The values of $\mu_{\text{h}}$ (solid, blue curve) and of $r_{\text{h}}/r_{\text{s}}$ (dashed, orange curve) as functions of the perturbation parameter $\beta$. The horizon parameter is chosen to be $\zeta_1=0$. Both describe the location of the putative horizon of the teleparallel black hole, and are related to each other through $r_{\text{h}}= \frac{r_{\text{s}}}{1-\mu_{\text{h}}^2}$. $\mu_{\text{h}}$ has been determined numerically.}
    \label{fig:murbetasingle}
\end{figure}

We show the plot of the location of the horizon in Figure~\ref{fig:murbetasingle} both in terms of $\mu$ and the more familiar $r$-coordinate, and display selected values of $\mu_{\text{h}}(\beta)$ and $r_{\text{h}}(\beta)$ in Table~\ref{tab:1}.

\begin{table}[h!]
    \centering
        \begin{tabular}{| c | c | c | c | c | c | c  | c | c |}
        \hline
        $\beta$                                            & 0.001  & 0.005  & 0.01   & 0.02   & 0.05   & 0.1    & 0.5    & 1 \\ \hline
        $\mu_{\text{h}}(\beta)$                            & 0.0698 & 0.1301 & 0.1664 & 0.2091 & 0.2744 & 0.3285 & 0.4591 & 0.5134 \\ \hline
        $\frac{r_{\text{h}}(\beta)}{r_{\text{s}}}$         & 1.0049 & 1.0172 & 1.0285 & 1.0457 & 1.0814 & 1.1210 & 1.2670 & 1.3580 \\ \hline
    \end{tabular}
    \caption{Numerical values of the event horizon radius in units of the Schwarzschild radius for selected values of $\beta$}.
    \label{tab:1}
\end{table}

It is clearly visible that the event horizon radius of a black hole increases for positive $\beta\leq 1$. Larger values of $\beta$ would definitely break the assumptions in arriving at the perturbative, dynamical equations of weak $f(\mathbb{T})$-gravity, and are therefore omitted at all times. We will later see that problems are to be expected even before $\beta=1$.

\subsection{Surface gravity and Black Hole Temperature}\label{ssec:BHtemp}

One essential quantity to analyze classical and semi-classical properties of black holes is the surface gravity, especially for thermodynamic aspects. Intuitively, the surface gravity $\kappa$ is given by the force at spatial infinity which is necessary to pull a massive point particle away from the black hole \cite{Poisson}. For a static, spherically symmetric metric, it is given by
\begin{align}
    \kappa = \frac{1}{2}\lim_{r\to r_{\text{h}}} \tfrac{A'(r)}{\sqrt{A(r)B(r)}}\,,
\end{align}
where $'$ denotes a derivative w.r.t. $r$. To study the properties of the black hole related to $\kappa$ it is essential that the geometry of spacetime under investigation satisfies $0<\lim_{r\to r_{\text{h}}} A(r)B(r)< \infty$, which we guaranteed by construction \textit{a forteriori}. We determined the metric components~\eqref{afinX2BH} and~\eqref{bfinX2BH} such that they satisfy $\lim_{r\to r_{\text{h}}} A(r)B(r)=1 + \beta \zeta_1$ for small $\beta \zeta_1$. As discussed at the end of section~\ref{ssec:ssym}, this property is not shared by the solutions found earlier in \cite{DeBenedictis:2016aze,Bahamonde:2019zea}, which is why it is not possible to study near horizon properties of the black hole in these solutions on the level of perturbation theory.

\begin{figure}[h!]
\centering
\begin{subfigure}[t]{.4\textwidth}
    \includegraphics[width=\textwidth]{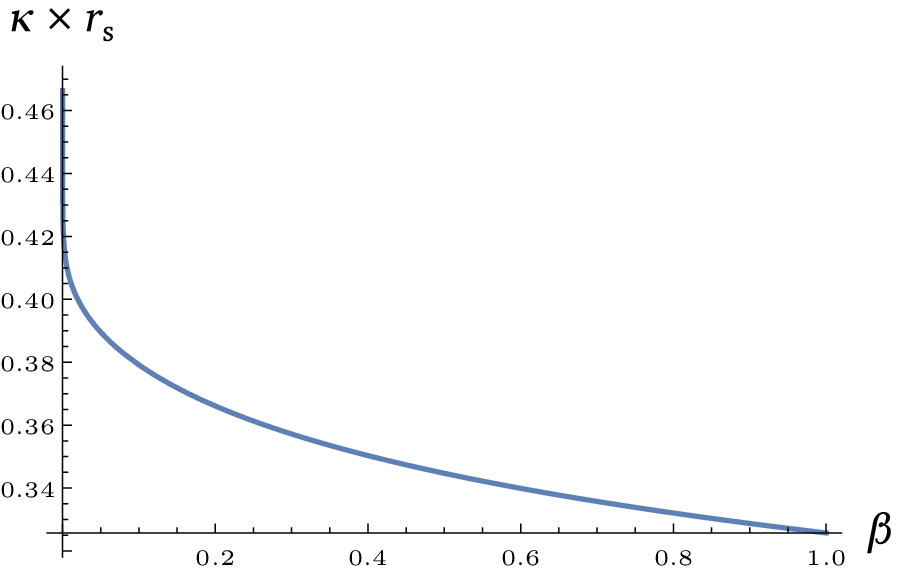}
    \caption{}
    \label{fig:sgvalues}    
\end{subfigure}
\begin{subfigure}[t]{.4\textwidth}
    \includegraphics[width=\textwidth]{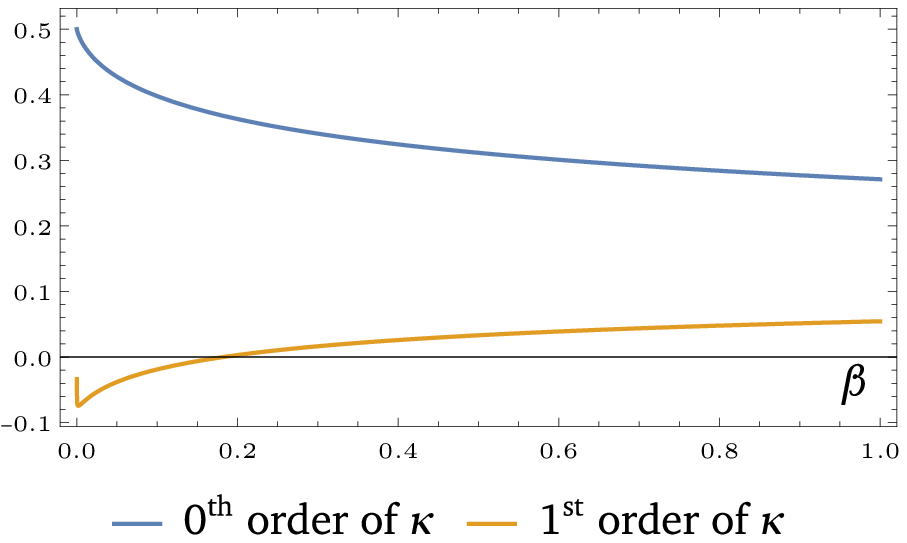}
    \caption{}
    \label{fig:sgterms}    
\end{subfigure}
\caption{\textit{Left:} Surface gravity in units of the inverse Schwarzschild radius for different values of the perturbation parameter $\beta$ and $\zeta_1=0$. The position needed for the evaluation of the surface gravity has been determined numerically. \textit{Right:} the values of the zeroth ($\kappa(r)|_{\beta=0,r=r_{\text{h}}}$) and first ($\beta (\partial_\beta\kappa(r)|_{\beta=0,r=r_{\text{h}}})$) order of $\kappa$~\eqref{eq:kappamuterms} plotted separately for a heuristic check of possible numerical issues.}
\label{fig:sg}
\end{figure}

To evaluate the surface gravity, we consider the quantity $\kappa(r) = \tfrac{A'(r)}{\sqrt{A(r)B(r)}}$. Its expansion to first order in $\beta$ can be obtained, using the ansatz~\eqref{eq:pertans} and the zeroth order terms from~\eqref{afinX2BH} and~\eqref{bfinX2BH}, as
\begin{align}
    \kappa(r) 
    &= \tfrac{a_0'(r)}{\sqrt{a_0(r)b_0(r)}} + \beta \frac{(2 a_0(r) b_0(r) a_1'(r) - a_0'(r) ( a_0(r) b_1(r) - a_1(r) b_0(r)  ) )}{\sqrt{a_0(r)b_0(r)}}, \\
    &= \frac{(1-\mu(r)^2)^2}{r_{\text{s}}} + \beta 2 \left( 2 a_1'(r) - \frac{(1-\mu(r)^2)^2}{r_{\text{s}}} \left[ \mu(r)^2 b_1(r) - a_1(r) \mu(r)^{-2}  \right] \right)\,.
\end{align}
To finally obtain the surface gravity, we evaluate $\kappa(r)$ at $r=r_{\text{h}}$. In terms of the variable $\mu = \sqrt{1-\frac{r_{\text{s}}}{r}}$ we find, also extracting the first order from~\eqref{afinX2BH} and~\eqref{afinX2BH},
\begin{align}
    \kappa = \frac{(\mu_{\text{h}}^2-1)^2}{2 r_{\text{s}}} \left( 1 - \frac{\beta}{6 \mu_{\text{h}}^2}\bigg[6 - 24 \mu_{\text{h}} + 31 \mu_{\text{h}}^2 + 24 \mu_{\text{h}}^3 - 42 \mu_{\text{h}}^4 - 40 \mu_{\text{h}}^5\right. \nonumber \\
        \left.\vphantom{\frac{\beta}{6 \mu_{\text{h}}^2}}  + 69 \mu_{\text{h}}^6 - 24 \mu_{\text{h}}^7 + 12 \mu_{\text{h}}^2  \ln(\mu_{\text{h}}) - 3 \mu_{\text{h}}^2 \zeta_1\bigg]  \right)\,.\label{eq:kappamuterms}
\end{align}
For increasing $\beta$, the surface gravity continuously decreases, and a plot of this behaviour is shown in Figure~\ref{fig:sg}. The second plot presented therein, Figure~\ref{fig:sgterms}, is meant as a heuristic to investigate convergence issues. As we will see later in our discussion of sparsity, this comparison can indicate when the numerically calculated, $\beta$-dependent horizon radius $r_{\text{h}}$ can impact how far we can trust the $\beta$-dependence of these \enquote{not-quite first-order in $\beta$ quantities}.

As figure~\ref{fig:sgvalues} shows, the surface gravity drops at very small values of $\beta$ very rapidly. However, it is possible to show that the limit $\beta\to 0$ is, in fact, well-defined and approaches the general relativistic value $\kappa_{\text{GR}} = \frac{1}{2 r_{\text{s}}}$. After this initial, sharp drop, the decline remains monotonous with $\beta$, but does not approach $0$ for values of $\beta$ befitting a perturbative interpretation.

The surface gravity of a black hole is directly related to the black hole temperature measured by an observer at infinity in the Hawking effect. This relation is simply
\begin{equation}
    T = \frac{\kappa}{2\pi},
\end{equation}
see for example \cite{Wald}, and has a corresponding thermal wave length:
\begin{align}
	\lambda_{\text{thermal}} = \frac{2\pi}{T}  
	=  \frac{8 \pi^2 r_{\text{s}}}{(\mu_{\text{h}}^2-1)^2} \Bigg( 1 + \frac{\beta}{6 \mu_{\text{h}}^2} 
	\bigg[&6 - 24 \mu_{\text{h}} + 31 \mu_{\text{h}}^2 + 24 \mu_{\text{h}}^3 - 42 \mu_{\text{h}}^4 - 40 \mu_{\text{h}}^5 \nonumber\\
	&+ 69 \mu_{\text{h}}^6 - 24 \mu_{\text{h}}^7 + 12 \mu_{\text{h}}^2  \ln(\mu_{\text{h}})- 3 \mu_{\text{h}}^2 \zeta_1\bigg]  \Bigg)\,.
\end{align}

With these quantities in place, we can have a first phenomenological look at the Hawking radiation in the next section using the concept of sparsity \cite{HawkFlux1,SparsityNdim}.

\subsection{Sparsity}\label{ssec:spars}
The concept of sparsity was introduced in \cite{HawkFlux1} to highlight and quantify an often overlooked feature of the Hawking effect: Between subsequent emission events of particles from the black hole significantly more time elapses than the localization time scale (roughly given by the frequency) of the emitted particles. Sparsity is the ratio between these two timescales---high sparsity means that many localization time \enquote{units} elapse before the next particle is emitted. Viewed in terms of length scales, this corresponds to the situation where the wavelength of the emitted particle is larger than the radius of the black hole. 

This breaks the usual analogy between the radiation of evaporating black holes with that of black bodies---or, accounting for graybody factors in the Hawking effect, graybodies. Usually, Hawking radiation is directly compared to the radiation of a blackbody. However, unlike black bodies encountered in thermodynamics where the emitted radiation is on average of a much shorter wavelength than the size of the blackbody itself, in black hole physics the emitter is of a smaller size than the emitted wavelengths. This is one of the different ways to recognize the relevance of sparsity. The lower the sparsity, the higher the density of states, the more classical the radiation, the more comparable to the radiation of a blackbody.

While this argument is best given using quantum theoretic arguments \cite{BHradNonclass}, heuristic arguments comparing the wavelength of the emitted radiation with the length scale of the emitter give a surprisingly accurate picture. While variations of this argument can be traced back at least to the work of Page in the second half of the 1970's \cite{Page1,Page2,Page3,PageThesis}, the reformulation in terms of sparsity has proved particularly fruitful for phenomenological studies of black holes both in GR and its various modifications \cite{HodNdim,SparsityBackreaction,OngGUPSparsity,SparsityNumerical,SparsityAna,SparsityScalarHair,SparsityAna2,SparsityNdim}.

Concretely, \enquote{sparsity} $\eta$ is defined as the ratio of above-mentioned time between emissions $\tau_{\text{gap}}$, and localization timescale $\tau_{\text{loc}}$:
\begin{equation}
    \eta = \frac{\tau_{\text{gap}}}{\tau_{\text{loc}}}.\label{eq:etadef}
\end{equation}
As is described in detail elsewhere \cite{HawkFlux1,SparsityNdim}, many different localization timescales present themselves for a given spectrum. In the present context, however, the precise choices are of less importance than the fact that for the Schwarzschild geometry, many different choices allow exact results, ranging from \numrange{28.4}{81.8}; values significantly larger than \num{1}.\footnote{In quoting these results, the multiplicity of 2 for massless particles has been taken into account.} For the Schwarzschild black hole, the only physical parameter, the mass $M$, conveniently cancels in these final results. In contrast, for actual blackbody radiation one has that $\eta\ll 1$.

\subsubsection{\texorpdfstring{Comparing weak $f(\mathbb{T})$-black holes and a Schwarzschild black hole}{Comparing weak f(T)-black holes and a Schwarzschild black hole}}

Rather than calculating the sparsity for the present, perturbative black hole spacetime in weak $f(\mathbb{T})$ gravity from scratch, the fact that we can regain the Schwarzschild spacetime as a limit for $\beta\to 0$ suggests to just work with the ratio of these two sparsities. We therefore assume that the Hawking spectrum still has the same blackbody nature (ignoring graybody factors), and look at the ratio ${\eta_{\text{weak }f(\mathbb{T})}}/{\eta_{\text{GR}}}$. By construction, the mass parameter $M$ entering through the Schwarzschild radius $r_{\text{s}}$ will also cancel for the sparsities of static, spherical symmetric black holes as they have been introduced above. As we work in the same spacetime dimensions, physical constants will likewise cancel in this ratio. Even more importantly, the cumbersome integrals needed for calculating sparsity will be the same, as both metrics are diagonal, static, spherical symmetric. In this ratio they therefore also drop out and we are left with an expression of the form
\begin{equation}
    \frac{\eta_{\text{weak }f(\mathbb{T})}}{\eta_{\text{GR}}} = \frac{\displaystyle\frac{\lambda_{\text{thermal, }f(\mathbb{T})}}{c_{\text{eff, }f(\mathbb{T})}\, A_{\text{h}, f(\mathbb{T})}}}{\displaystyle\frac{\lambda_{\text{thermal, GR}}}{c_{\text{eff, GR}}\, A_{\text{h}, GR}}},
\end{equation}
where $\lambda_{\text{thermal}}$ is the thermal wavelength of the corresponding black hole solution in weak $f(\mathbb{T})$-gravity and GR, respectively, and $A_{\text{h}} = 4\pi r_{\text{h}}^2$ is the horizon area of each solution. The quantity $c_{\text{eff}}$ requires some more discussion, harking back to comments made in the beginning of this section: A thermodynamic blackbody has a size that is large compared to the wavelength of its radiation. To highlight the difference between a black hole and a blackbody as strongly as possible, sparsity was constructed as conservative a measure as possible: By potentially underestimating the sparsity, bringing it closer to the situation encountered in a blackbody, any result for which $\eta\gg 1$ becomes stronger. Therefore, the area from which the radiation is emitted was not chosen to be the horizon but a larger surface. 

While perhaps counter-intuitive for people who like to think of Hawking radiation as a tunneling process, there are good reasons to do this: First of all, interpreting Hawking radiation as a tunneling process just through the horizon is difficult to defend in light of $\lambda_{\text{thermal}}\gg r_{\text{s}}$.\footnote{An $s$-wave derivation of the Hawking effect as in \cite{ParikhWilczekTunnelling} nevertheless has its place and pedagogical value, and the tunneling interpretation many helpful insights. One just needs to be extremely careful how far to trust its implications.} Instead of $A_{\text{h}}$ the surface of the emitter was chosen in \cite{HawkFlux1} to be the capture cross-section $A_{\text{cap}} = \pi b_{\text{min}}^2 r_{\text{h}}$. This is (in part) motivated by results on renormalized stress-energy tensors in black hole spacetimes \cite{BHQuantumAtmosphere}: There it was found that energy density and fluxes peak much further out than at the horizon. If we now take a sphere of a surface area $A_{\text{cap}}$, this is much closer to this, then the horizon area would be. As the capture cross-section also enters the calculation of graybody factors, it also immediately allows easier comparison with the literature when graybody factors are included \cite{Page1,Page2,Page3,PageThesis,FrolovZelnikov2011,SparsityNumerical}. We now define $c_{\text{eff}}$ as
\begin{equation}
    c_{\text{eff}} = \frac{A_{\text{cap}}}{A_{\text{h}}}.
\end{equation}
For GR, we have that $c_{\text{eff, GR}} = 27/16$ \cite{FrolovZelnikov2011}. In our context of weak $f(\mathbb{T})$-gravity, the result is a much more cumbersome
\begin{align}
	c_{\text{eff}, f(\mathbb{T})} = \frac{27}{16} + \frac{2\sqrt{3}\beta}{16} \Bigg(-64-\frac{80}{\sqrt{3}}-\frac{\sqrt{3}}{\mu_{\text{h}}^2}\bigg[&-6+12 \mu_{\text{h}}-21\mu_{\text{h}}^2-108\mu_{\text{h}}^3+66\mu_{\text{h}}^4\\ &+20\mu_{\text{h}}^5-39\mu_{\text{h}}^6+12\mu_{\text{h}}^2(\log\mu_{\text{h}}+\mu_{\text{h}}^5)\bigg]\Bigg)\,.\nonumber
\end{align}
We followed the previous strategy in arriving at a \enquote{first order in $\beta$ quantity}.

\begin{figure}
    \centering
    \begin{subfigure}[t]{.45\textwidth}
        \vskip 1pt
        \includegraphics[width=\textwidth]{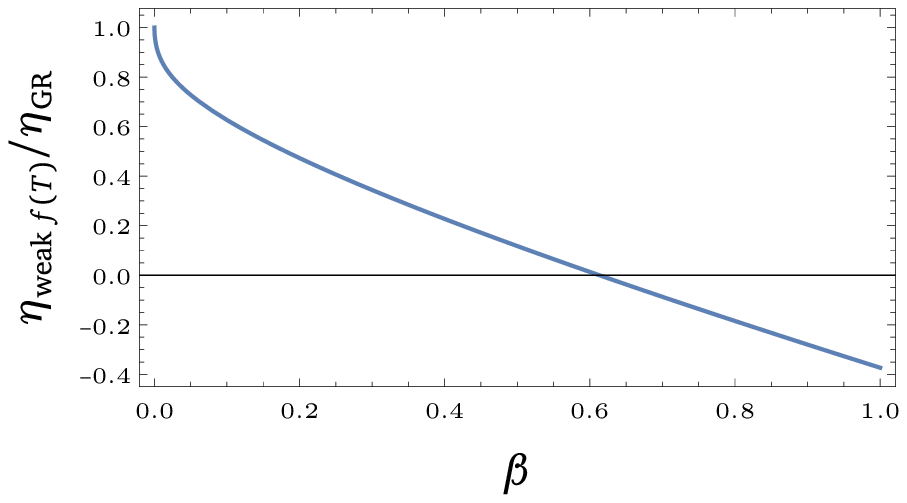}\\
        \vspace{1.3\baselineskip}
        \caption{}
        \label{fig:sparsityvalue}
    \end{subfigure}
	~~
    \begin{subfigure}[t]{.45\textwidth}
        \vskip 0pt
        \includegraphics[width=\textwidth]{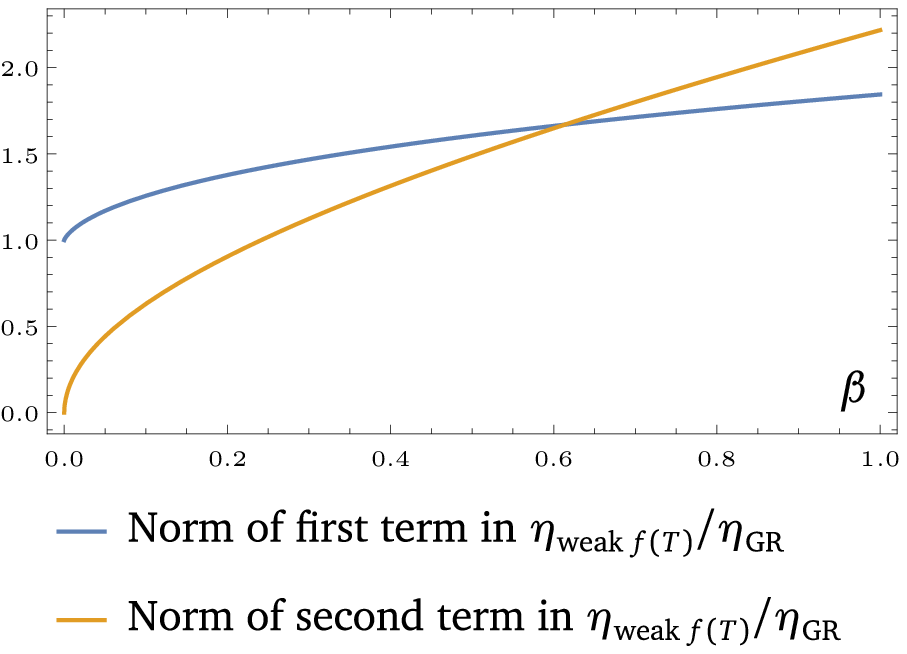}
        \caption{}
        \label{fig:sparsityterms}
    \end{subfigure}
    \caption{\textit{Left:} The sparsity of a weak $f(\mathbb{T})$ black hole compared to the sparsity of a Schwarzschild black hole as a function of $\beta$. \textit{Right:} A look at the terms of our perturbative approach to this ratio reveals a breakdown of this approach for sufficiently large $\beta$ somewhere below $\beta\approx\num{0.6}$.}
    \label{fig:sparsity}
\end{figure}

Putting everything together (in the by now standard way for handling the perturbative parameter $\beta$), we finally arrive at
\begin{align}
	\frac{\eta_{\text{weak }f(\mathbb{T})}}{\eta_{\text{GR}}} = \frac{1}{(\mu_{\text{h}}^2-1)^2} -\frac{2\beta}{9\mu_{\text{h}} (\mu_{\text{h}}^2-1)^2} 
	\bigg(&18-[55+32\sqrt{3} + 36\log{\mu_{\text{h}}}]\mu_{\text{h}}\\
	&+126 \mu_{\text{h}}^2 - 36 \mu_{\text{h}}^4+30 \mu_{\text{h}}^4 - 45\mu_{\text{h}}^5+18\mu_{\text{h}}^6\bigg).\nonumber
\end{align}
The results for sparsity as compared to the Schwarzschild case are shown in figure~\ref{fig:sparsity}. Looking at the values of this ratio in figure~\ref{fig:sparsityvalue}, however, clearly shows that somewhere before $\beta\approx\num{0.6}$ our attempt at a perturbative treatment breaks down. Since the temperature (through the surface gravity) is perfectly well-behaved, this is a new artefact of our approach at this stage. A look at the plot in~\ref{fig:sparsityterms} demonstrates that our heuristic of comparing the term \enquote{linear in $\beta$} with the \enquote{zeroth order} indeed picks up this issue. Hence, the sparsity results should only be trusted small values of $\beta\ll \num{0.6}$.

While the sparsity is monotonically decreasing with $\beta$, this does not happen drastically enough to turn the black hole radiation into classical blackbody radition. It will not be reduced enough to cancel the large sparsity values found for a Schwarzschild black hole (\numrange{28.4}{81.8}) and the radiation stays sparse. Sparsity will remain, unless our calculated quantity can be trusted even close to the point where the first and second term in figure~\ref{fig:sparsityterms} cross. This seems unlikely, and a weak $f(\mathbb{T})$-gravity black hole would hence not change the quantum nature of the Hawking radiation associated with it. The thermal wavelength remains larger than the horizon radius, the density of states of the radiation remains low.

\section{Conclusion}\label{sec:conc}
In this article we have identified the general black hole family of spherically symmetric, static vacuum solutions of weak $f(\mathbb{T})$-gravity in equations~\eqref{afinX2BH} and~\eqref{bfinX2BH}. The crucial difference to the spherically symmetric, static vacuum solutions which have been found earlier is that the choice of integration constants in these earlier results precludes an interpretation as a black hole on perturbative grounds. Our choice of integration constants ensures that this interpretation remains possible. Asymptotic flatness is not quite as obvious as in the earlier approach (where it was a key ingredient in the choice of integration constants), but after a simple redefinition of coordinate time easily regained.

The analysis of observable consequences of this teleparallel perturbation of Schwarzschild geometry revealed that the corrections manifest themselves strongest in a larger photon sphere, a reduced black hole temperature and a slightly lower, though qualitatively similar sparsity. The classical observables of the perihelion shift, the Shapiro delay and the deflection of light only acquire very small corrections in the weakly teleparallel black holes spacetimes, a behaviour which is different to the solutions found earlier. Here, the impact of the teleparallel perturbation on these observables is much higher \cite{DeBenedictis:2016aze,Bahamonde:2019zea}. In particular, our results also quantify the behaviour of the exterior of the BTZ black holes as found in Born--Infeld gravity \cite{Boehmer:2019uxv,Boehmer:2020hkn} for small parameter $\frac{1}{\lambda}$ (which is $\alpha$ in our notation). 

Hence, to constrain or discover teleparallel perturbations, our results indicate that the best observable are lensing observables of light rays passing the region close the horizon. The best observable of this kind is the shadow of black holes \cite{Akiyama:2019cqa}. As a first step, the black hole shadow of the spherical symmetric weak $f(\mathbb{T})$ black hole can be derived. To obtain the influence of the teleparallel perturbation on a realistic black hole shadow the whole derivation has to be extended to rotating axially symmetric black holes. The derivation of teleparallel perturbations of Kerr spacetime is currently work in progress, and first steps towards this goal have been achieved \cite{Bahamonde:2020snl}.

\begin{acknowledgments}
The authors would like to thank Sebastian Bahamonde and Matt Visser for valuable discussions. CP was funded by the Deutsche Forschungsgemeinschaft (DFG, German Research Foundation) - Project Number 420243324. SS was supported by OP RDE project No. CZ.02.2.69/0.0/0.0/18\_053/0016976 International mobility of research, technical and administrative staff at the Charles University, and also acknowledges partial and indirect support by the Marsden Fund of the Royal Society of New Zealand. 
\end{acknowledgments}

\bigskip

\appendix

\small
\bibliographystyle{cmphref}
\bibliography{fTbh.bib}

\end{document}